\documentclass[12pt]{article}
\usepackage[utf8]{inputenc}
\usepackage{amsmath}
\usepackage{amssymb}
\usepackage{graphicx}
\usepackage[a4paper, total={4.75in, 7.5in}]{geometry}
\usepackage[labelfont=bf]{caption}
\usepackage[english]{babel}
\addto\captionsenglish{}

\newenvironment{sciabstract}{%
\begin{quote} \bf}
{\end{quote}}

\usepackage{scicite}
\usepackage{times}
\usepackage{setspace}

\title{Twistons in a Sea of Magic}
\author
{Simon Turkel,$^{1}$ Joshua Swann,$^{1}$ Ziyan Zhu,$^{2}$ Maine Christos,$^{2}$\\ 
K. Watanabe,$^{3}$ T. Taniguchi,$^{4}$ Subir Sachdev,$^{2}$ Mathias S. Scheurer,$^{5}$\\ 
Efthimios Kaxiras,$^{2,6}$ Cory R. Dean,$^{1}$ and Abhay N. Pasupathy$^{1,7,*}$\\
\\
\normalsize{$^{1}$Department of Physics, Columbia University, New York, NY 10027, USA}\\
\normalsize{$^{2}$Department of Physics, Harvard University, Cambridge, MA 02138, USA}\\
\normalsize{$^{3}$Research Center for Functional Materials,}\\ 
\normalsize{National Institute for Materials Science, 1-1 Namiki, Tsukuba 305-0044, Japan}\\
\normalsize{$^{4}$International Center for Materials Nanoarchitectonics,}\\
\normalsize{National Institute for Materials Science, 1-1 Namiki, Tsukuba 305-0044, Japan}\\
\normalsize{$^{5}$Institute for Theoretical Physics, University of Innsbruck, Innsbruck A-6020, Austria}\\
\normalsize{$^{6}$John A. Paulson School of Engineering and Applied Sciences,}\\
\normalsize{Harvard University, Cambridge, MA 02138, USA} \\
\normalsize{$^{7}$Condensed Matter Physics and Materials Science Department,} \\
\normalsize{Brookhaven National Laboratory, Upton, NY 11973, USA}\\
\normalsize{$^*$ email address: apn2108@columbia.edu}\\
}

\date{}

\begin{document}
\doublespacing

\maketitle
\newpage

\begin{sciabstract}

Magic angle twisted trilayer graphene (TTG) has recently emerged as a new platform to engineer strongly correlated flat bands.  Here, we reveal the structural and electronic properties of TTG using low temperature scanning tunneling microscopy at twist angles for which superconductivity has been observed.  Real trilayer samples deviate from their idealized structure due to a strong reconstruction of the moir\'e lattice, which locks layers into near-magic angle, mirror symmetric domains comparable in size to the superconducting coherence length.  The price for this magic relaxation is the introduction of an array of localized twist angle faults, termed twistons. These novel, gate-tunable moir\'e defects offer a natural explanation for the superconducting dome observed in transport and provide an avenue to probe superconducting pairing mechanisms through disorder tuning.

\end{sciabstract}

The prediction of a magic angle in twisted trilayer graphene (TTG) \cite{magichierarchy,1907.12338}, at which electrons lose their kinetic energy, laid out a promising pathway to extend the moir\'e paradigm into three dimensions.  This was soon followed by measurements of superconductivity and field dependent quantum interference \cite{pjh,kim,triplet}, making TTG the only moir\'e heterostructure outside of magic angle twisted bilayer graphene (MATBG) to exhibit signatures of both a superconducting transition and macroscopic quantum phase coherence.  It has been proposed, since both TTG and MATBG share the unique attribute of two-fold rotational symmetry $C_{2z}$, that this symmetry is essential to establishing superconductivity in twisted graphenes \cite{kim,Yazdani1098}.  In fact, superconductivity in TTG appears to be even more robust than in MATBG, with $\text{T}_\text{c}$ reaching up to 2.9K in the first generation of devices.  This has led to speculation that magic angle TTG is structurally more stable than MATBG, locking experimental devices into a mirror symmetric configuration that possesses the crucial $C_{2z}$ symmetry.  Theoretical works have proposed several exotic orders for the mirror symmetric configuration, including spontaneous flavor-symmetry breaking, nematic superconductivity, and spin triplet pairing \cite{Christos2021,Fischer2021,macd_triplet}.  To date, however, virtually nothing is known from experiment about the atomic or electronic structure of this novel material, so that theoretical studies are largely based upon simple models that may or may not capture the intrinsic physics of real devices.  Indeed, there remains no direct experimental confirmation of even the most basic hypothesis, that superconducting devices possess the mirror symmetric stacking upon which theoretical predictions are based.

TTG is formed by consecutively stacking three layers of graphene so that the bottom layer (B) is rotated at an angle $\theta_{BM}$ relative to the middle layer (M) and the top layer (T) is rotated at an angle $\theta_{TM}$ relative to the middle layer (Fig.~1A, inset).  Each rotation $\theta_{ij}$ gives rise to a periodic density modulation, or moir\'e pattern, at wavelength $\lambda_{ij}\sim a/\theta_{ij}$, where $a=0.246~\textrm{nm}$ is the graphene lattice constant \cite{Yankowitz2012,Li2010,Wong2015}.  For the special case of mirror symmetric stacking, $\theta_{BM} = \theta_{TM}=\theta$ (i.e. T and B are aligned and M is twisted relative to these by an angle $\theta$), TTG is predicted to host two sets of flat bands whose band velocity vanishes at a magic angle of $\theta\sim1.56^\circ$ \cite{magichierarchy,1907.12338}.  As in MATBG, the quenched kinetic energy of charge carriers in these bands is expected to favor the formation of strongly correlated states of matter.  Indeed, recent transport measurements have confirmed the importance of electronic correlations in TTG with the observation of superconductivity by two groups with similar phenomenology \cite{pjh,kim}.  

Several obstacles can stand in the way of achieving perfect mirror symmetry.  Despite state-of-the-art fabrication techniques, the highest quality TTG heterostructures will inevitably have a small mismatch between $\theta_{TM}$ and $\theta_{BM}$ over macroscopic length scales, as was the case in at least one superconducting device \cite{kim}.  In the limit of perfectly rigid graphene layers (i.e., neglecting lattice relaxation), such a misalignment will produce a beating pattern between the top-middle (TM) and bottom-middle (BM) moir\'es at \textquotedblleft{}moir\'e of moir\'e\textquotedblright{} wavelength $\Lambda\sim a/\delta_\theta$, where $\delta_\theta = |\theta_{TM}-\theta_{BM}|$ (Fig.~1A).  In regions where the two moir\'es are in phase, TM AA sites sit atop BM AA sites, resulting in a locally mirror symmetric AtA ({\textquotedblleft}A-twist-A{\textquotedblright}) trilayer configuration comprised of AAA, ABA, and BAB stacking sites.  Where the two moir\'es are out of phase, by contrast, the AA sites of one bilayer align with the AB sites of the other, generating a local AtB configuration \cite{macD, wu2020}, comprised of ABB, AAB, and BAC stacking sites, which is related to the mirror symmetric configuration by translation of the top layer (Figs.~1B and 1C).  The emergent structures of the trilayer moir\'es in these two regions are  distinguished by their different symmetry classes, as visualized by their predicted topographic profiles in Fig. 1B.  We estimate the out-of-plane corrugations for AtA and AtB domains as a superposition of sinusoidal functions of local bilayer stackings with maxima on AA and minima on AB sites \cite{koshino_maxloc}; we find that whereas the AtB regions host a honeycomb moir\'e lattice, the moir\'e pattern in the AtA domains is expected to be hexagonal.

In this work, we utilize the atomic scale imaging capabilities of ultra-high vacuum scanning tunneling microscopy and spectroscopy (STM/S) at temperatures from 4.8K to 7.2K to directly characterize the electronic structure of magic angle TTG.  Our devices are fabricated using the {\textquotedblleft}cut and stack{\textquotedblright} technique and electrical contact is made with a pre-placed graphite finger to which Field's metal $\mu$-solder is subsequently affixed (Fig.~S1).  Fig.~1D shows STM topography of a TTG sample in which two distinct moir\'e wavelengths, $\lambda \sim 9$ nm and $\Lambda \sim 70$ nm, are clearly visible, corresponding to the bilayer moir\'e and moir\'e of moir\'e length scales respectively.  The corresponding angle mismatch $\delta_\theta\sim a/\Lambda$ for this region is $\sim0.2^\circ$, which is nearly identical to the mismatch of $\sim0.3^\circ$ measured in a superconducting TTG device \cite{kim}.  The presence of two moir\'e patterns is a generic feature over large areas of our sample (Fig.~S2) and represents a deviation from the three moir\'es ($\lambda_{TM}, \lambda_{BM},$ and $\Lambda$) that are expected on the basis of a simple rigid model.  Fig.~1E shows a zoomed in topograph of a single unit cell of the larger moir\'e periodicity in this region.  The STM signal in constant current mode is dominated by structural height variations across the sample surface (Fig.~S3), so that we can identify the global stacking configuration as AtA by the fact that the smaller moir\'e lattice is hexagonal rather than honeycomb at each point in space.  The bright spots in topography therefore correspond to regions of local AAA stacking and are surrounded by alternating ABA and BAB domains, which is confirmed by line cut spectroscopy (Fig.~S4).

The absence of AtB domains in a sample with nonzero angle mismatch $\delta_{\theta}$ implies that TTG undergoes a reconstruction on the scale of the moir\'e lattice that favors the lower energy \cite{energybarrier} AtA configuration.  Close examination of Fig.~1E reveals that this moir\'e lattice reconstruction (MLR) produces a periodic warping  of the AAA site positions in order to enforce AtA stacking over the entire sample area.  The observed warping of the moir\'e lattice can be understood at the atomic scale as arising from variations in the local twist angle ($\theta_x$) and strain ($\varepsilon_x$) of the individual graphene layers.  Figs.~1F and 1G plot $\theta_x\sim a/\sqrt{A_x}$,  for two nearby sample regions, where $A_x$ is the area of the moir\'e unit cell centered on position $x$.  For small angle mismatch $\delta_\theta$, the system segregates into highly uniform triangular domains (blue areas) bounded by sharp point-like irregularities in the local twist angle (red areas).

Remarkably, for $\Lambda\gtrsim 30$ nm, the average twist angle within each domain ($\theta_I$) saturates to a common value of $\sim1.5^\circ$ that is independent of the relative orientation of the top and bottom layers (Fig.~1H), which agrees with theoretical structural relaxation calculations (Fig.~S6).  This suggests a natural tendency for TTG to locally lock to the mirror symmetric magic angle structure while {\textquotedblleft}absorbing{\textquotedblright} twist angle inhomogeneity at the larger moir\'e of moir\'e length scale.  This does not, however, come at zero cost.  The effect of the MLR on the local electronic structure is profound, as evidenced by the Fermi level local density of states (LDOS) map presented in Fig.~1I, which shows large modulations in the tunneling conductivity across regions of the MLR.  It is therefore necessary, in considering the potentialities of TTG as a platform for correlated phases, to analyze the electronic structure on both the sub- and supra-$\Lambda$ length scales, which we now proceed to do in turn.

In Fig.~2A we present STM topography of a $250\text{ nm}^2$ area, which is part of an even larger region with only a single moir\'e wavelength corresponding to a twist angle of $\theta = 1.55^\circ$. The extreme degree of homogeneity in this area is conveyed by the local twist angle histogram (inset), showing a standard deviation of $0.03^\circ$ over the entire field of view.  This indicates a twist angle mismatch of $\delta_{\theta}<0.05^\circ$, providing us with the opportunity to study a single domain of the MLR, as well as to investigate the spectroscopic properties of a large patch of magic angle TTG that is approaching the size of a transport device.

The high energy resolution of STS permits us to directly probe the  structure of the flat bands.  Fig.~2B displays a series of STS measurements acquired at 7.2K on a single AAA site for a range of voltages ($V_\text{g}$) applied to the graphite back gate.  The measured spectrum does not change appreciably upon cooling to 4.8K (Fig.~S7).  At charge neutrality (CNP) the spectrum is dominated by a pair of overlapping resonances, arising from the partially overlapped conduction (CB) and valence (VB) flat bands.  Additional soft humps at higher energy (black arrows) correspond to the edges of the next available (remote) bands.  Each flat band is expected to host a saddle point in its momentum space structure, giving rise to a sharp peak, or van Hove singularity (VHS), in the density of states.  We extract the energy positions and widths of these VHSs by fitting our spectra with the sum of two Lorentzian curves and find that at CNP the CB and VB VHSs are separated by $\sim$18~meV and have an average width (FWHM) of $\sim23$~meV.  

Varying $V_\text{g}$ systematically alters the shape of the quasiparticle spectrum, changing the intensities, separations, and widths of the flat band VHSs.  In particular, we find a transfer of spectral weight between the two VHSs upon reversing the sign of $V_\text{g}$ (Fig.~2B).  Moreover, the width of each flat band is reduced when it is doped to the Fermi level, saturating to a minimum width of $\sim15$~meV at $\nu\sim\pm2$ (Fig.~2C).  Lastly, as shown in Fig.~2D, the VHS separation is an increasing function of doping away from CNP with a distinct asymmetry between filling of electrons and holes.  In general, such gate dependent spectral shifts can be attributed to either the single particle effect of displacement field ($D=V_\text{g}/2d$) on the material's band structure, or to variations in the quasiparticle interaction strength as a function of band filling ($\nu=4n/n_s$), where $d$ is the dielectric thickness, $n$ is the induced carrier density, and $n_s$ is the carrier density at full filling of a fourfold degenerate moir\'e band.  

We can examine the role of interactions in determining the band structure of TTG by comparing the experimental spectrum with continuum model \cite{BMD,1907.12338,magichierarchy} calculations for a uniform mirror symmetric AtA stacking configuration at $\theta=1.55^\circ$.  Figs.~2E and S8A compare the measured VHS separation and widths at CNP with those predicted by three separate calculations.  Using inter- and intra-layer tunneling parameters (see SM) derived from \textit{ab initio} computations \cite{macD} severely underestimates both the separation and widths of the VHSs (SP1).  Enhancing the monolayer graphene Fermi velocity by $\sim30\%$ (SP2) enables us to reproduce the VHS separation, but predicts widths that are still a factor of $\sim6$ less than those found in experiment.  As a doping dependent calculation including electron interactions \cite{Christos2021} is beyond the scope of this primarily experimental work, we here restrict the theoretical analysis of interactions to the CNP. Employing the self-consistent Hartree-Fock procedure of Ref.~\cite{Christos2021} (see SM for more details), we find, similar to the situation in MATBG \cite{alex,Xie2019}, that the interaction-induced band renormalization without additional symmetry breaking accurately captures both the separation between the peaks and their broadening; to reproduce the width quantitatively, an additional lifetime broadening of $4\,\textrm{meV}$ is sufficient (Figs.~2E and S8A).

As noted above, tuning $V_\text{g}$ away from zero produces systematic changes in the intensities, separation, and widths of the VHSs.  In Fig.~S8B we examine the effect of $D$ on the single particle band structure and find that while it does account for the shift in relative intensity of the VHSs, values of $D$ within the experimentally accessible range fail to produce notable changes in either the predicted separation or widths of the VHSs.  The inability of single particle calculations to reproduce the measured quasiparticle spectrum, combined with the strong doping dependence of the latter, provides clear evidence for a pronounced band renormalization in TTG near the magic angle due to strong quasiparticle interactions.

In order to isolate and better visualize this reconfiguration of the band structure, we plot in Fig.~2F gate dependent STS with each curve shifted so that the flat bands remain centered on zero energy.  The position of the chemical potential at each doping is denoted by the red dashed line.  Unlike MATBG, in which superconductivity occurs at ubiquitous filling factors of the moir\'e bands \cite{Lu2019}, observations of superconductivity in TTG have been strictly limited to within the vicinity of $|\nu|$ = 2, with optimal doping occurring for $2<|\nu|<3$ \cite{pjh,kim}.  In MATBG superconductivity occurs when the chemical potential is embedded in the moir\'e flat bands, leading to a large density of states at the Fermi level.  In TTG, we can likewise ask whether there are any spectroscopically unique features that emerge in the vicinity of the parent states of superconductivity, marked with pink arrows in Fig.~2F. 

The inherent electron-hole asymmetry of our spectrum implies that the hole- and electron-doped parent states correspond to highly distinct Fermi surfaces.  Whereas the chemical potential lies roughly within the VB VHS in the hole doped parent state, it sits $\sim10$~meV above the CB VHS in the electron doped parent state.  The resulting enhancement of the Fermi level density of states for hole doping may explain the comparative robustness of the hole doped superconducting dome observed in transport, however it also implies that there is no clear correlation between the precise Fermi surface of the parent state and where in the phase diagram superconductivity has been observed.  A simple hypothesis to alleviate this apparent discrepancy, and to which we appeal below, is that the MLR is instrumental in determining the boundaries of the superconducting phase.

Having analyzed the electronic structure at the sub-$\Lambda$ length scale, we now turn to a detailed study of the MLR at twist angles near those for which robust superconductivity has been observed \cite{kim,pjh}.  Fig.~3A shows a large area topograph of a sample region with angle mismatch $\delta_{\theta}\sim0.25^\circ$.  Overlaid on the topography of Fig.~3A is a map of the local twist angle, giving a spatially averaged value of $\bar{\theta}=1.55^\circ$.  As we have seen, the MLR segregates the system into domains of uniform \textit{magic moir\'e} arranged in a honeycomb lattice with $\theta_x\sim1.5^\circ$ separated by quasi-one-dimensional \textit{moir\'e solitons}.  Populating the nodes of this soliton network is a hexagonal lattice of point-like faults in the local twist angle corresponding to topological moir\'e defects that we term \textit{twistons}.  Zoomed in topographs of these three regions of the MLR are shown in Figs.~3B-D.  Of the three sites, only the moir\'e solitons show considerable breaking of $C_3$ rotational symmetry, which is consistent with a uniquely large value of local heterostrain $\varepsilon_x\gtrsim0.5\%$ on these structures \cite{Kazmierczak2021}.

The large variation in $\theta_x$ and $\varepsilon_x$ on the $\Lambda$ scale has a dramatic effect on the local electronic structure of TTG.  $\theta_x$ serves as a convenient parameter to quantitatively classify different regions of the larger moir\'e, since each region roughly corresponds to a unique value.  Fig.~3E shows LDOS spectra acquired on AAA sites as a function of increasing $\theta_x$.  Alongside these, in Fig.~3F, we plot continuum model (SP2) densities of states for a series of structural parameters ($\theta,\varepsilon$) approximating those found in the respective experimental topography.  For low relative twists, corresponding to the magic moir\'e, the spectrum approximates that expected for TTG with a uniform $\theta\sim1.45^\circ$.  As we increase $\theta_x$, moving onto the moir\'e solitons, the spectral intensity of the flat bands is progressively attenuated to the point of being practically indistinct, as expected for a highly strained TTG system.  Our calculation assumes a uniaxial strain applied only to the middle layer \cite{heterostrain,tdbg}, which likely underestimates the experimental effect of $\varepsilon$, for which strain is distributed in a non-uniform way throughout all three layers.  Increasing $\theta_x$ still further (hence decreasing $\varepsilon_x$), we find that the flat bands regain their intensity, but are now split apart in energy by approximately 40~meV, consistent with our calculation for unstrained TTG at $\theta\sim1.8^\circ$.  These observations indicate that the local electronic structure of TTG at small but finite $\delta_\theta$ is primarily determined by the local values of heterostrain and twist angle given by the MLR.  Twist angle disorder in TTG does not, therefore, result in a smooth and continuous fluctuation of the electronic structure, as it does in MATBG \cite{Zondiner2020}, but rather leads to the formation of electronically isolated grains, yielding an inherently inhomogeneous flat band system.

The expectation that small twist angle disorder in TTG leads to a spontaneous segregation into domains with distinct local twist angles, and hence distinct electronic structures, is borne out by structural relaxation calculations \cite{zoerelaxation}.  Fig.~3G shows the results of such a calculation for $\theta_{TM} = 1.5^\circ$ and $\theta_{BM} = 1.69^\circ$, where the local twist angle has been extracted from the average distance between nearest neighbor AAA sites.  As in Fig.~3A, the sample assumes universal AtA stacking (see Fig.~S5E) by separating into three regions characterized by distinct values of local twist angle.  This is illustrated by the corresponding histogram in Fig.~3H.  While this calculation confirms the qualitative features of the experimental topography, it does not generate a quantitative match, possibly due to the effect of out of plane corrugations, which are not explicitly considered in the model, or due to the neglected interaction between top and bottom layers. 

We can explore the implications of this structural and electronic inhomogeneity for the correlated states at partial fillings by performing STS measurements as a function of $V_\text{g}$.  Figs.~4A and 4B show characteristic filling dependent spectroscopy measured on AAA sites of the magic moir\'e and twiston respectively.  Full filling of the moir\'e superlattice can be identified as the carrier density at which the derivative of the chemical potential, $d\mu/dn$, undergoes a rapid step-like increase, indicated by the yellow arrow in Fig.~4A.  In TTG each moir\'e band is fourfold degenerate, so that full filling corresponds to a density $n_s = 4/A$, where $A$ is the moir\'e unit cell area \cite{kim,pjh,Cao2018}.  In our case, the size of the moir\'e unit cell is a function of position in the MLR ($A\rightarrow A(x)$), so that we must refer to a local filling factor $\nu_x = nA(x)$.  In order to facilitate comparisons between our local measurements and the phase diagram gleaned from bulk probe assays, we provide in Fig.~S9 a chart of the statistical prevalence of local filling factors as a function of induced carrier density.  The average value, $\bar{\nu}$, is an approximation of the quantity probed in transport. 

In spectroscopic measurements, correlation induced states typically appear as spectral gaps centered on the Fermi level that emerge and disappear as a function of induced carrier density \cite{alex,Jiang2019,Xie2019,Choi2019,crommieTDBG}.  In our measurements, we do not observe spectral gaps in regions of uniform magic moir\'e (Figs.~2B, 4A, and S7), consistent with the predominantly metallic behavior measured in transport.  At low temperature, however, the longitudinal resistivity has been shown to develop a non-trivial doping dependence, with local maxima emerging at $\nu\sim0$ and $\nu\sim\pm2$ \cite{pjh,kim}.  These interaction induced resistive states (IIR) are not expected from the single particle band structure and are reminiscent of the correlated insulators at integer fillings of MATBG.  Examining the doping dependent twiston spectrum (Fig.~4B), we find that there is a pronounced reduction of the LDOS centered about the Fermi level at $\bar{\nu}\sim0$ and $\bar{\nu}\sim\pm2$ (red arrows), presenting clear signatures of electronic correlations that coincide with the onset of IIR behavior.  The real space confinement of this resistive phase to the twiston sites implies that the interaction strength is strongly modulated on the $\Lambda$ length scale and offers an explanation for the weakly resistive behavior observed in transport.

We now reexamine the parent state out of which superconductivity emerges in the context of the observed MLR.  The differential rates of band filling on the regions of the $\Lambda$-modulation (Fig.~S9) mean that as we add charge to the system we are simultaneously tuning the twiston and magic moir\'e flat bands relative both to the chemical potential and to one another.  This is illustrated in Figs.~4C and 4D, which show calculated band fillings at two values of $n$ for twist angles of 1.45$^\circ$ and 1.8$^\circ$.  In Fig.~4E we overlay the flat band spectra on twiston and magic moir\'e sites for the full range of measured fillings.  There exists a small range of $n$ for which the two sets of flat bands are in approximate resonance with one another, giving rise to an enhanced Fermi level density of states, which favors electronic correlations.  In Fig.~4F we quantify this \textit{flat band resonance} by plotting the energy difference between spatially separated flat bands as a function of doping.  The resonance condition is satisfied for $2\lesssim|\bar{\nu}|\lesssim3$, in the precise region of optimal doping for superconductivity.  We expect the range of resonant dopings to be largely independent of the particular value of the twist-angle mismatch $\delta_{\theta}$ in a given sample due to the observed relaxation phenomenon described above (Figs.~1F-H and S6), so that the regime of optimal doping would be roughly constant across samples with $\delta_{\theta}\lesssim0.5^\circ$.  At $|\bar{\nu}|\gtrsim3$, the flat band resonance is abruptly destroyed as the magic moir\'e VHS is doped past full filling and swept rapidly away from the Fermi level, consistent with a sharp boundary to the superconducting state on the overdoped side. 

We can gain further insight into the nature of the parent state by examining the effect of the flat band resonance on the real space electronic structure through doping-dependent LDOS mapping.  Figs.~4G-I show LDOS maps acquired at the Fermi level for the three carrier densities marked by arrows in Fig.~4F (see Fig.~S10 for energy dependence).  As discussed above, the sample displays considerable disorder at charge neutrality (Fig.~4H).  The angle mismatch $\delta_\theta$ in this region, as in superconducting devices \cite{kim}, is $\sim0.3^\circ$, leading to magic moir\'e grains of lateral dimension $\sim50$~nm, similar in magnitude to the superconducting coherence length \cite{pjh,kim}.  As we tune the carrier density towards the flat band resonance, however, the LDOS maps become increasingly homogeneous (Figs.~4G and 4I), indicating a reduction in the strength of the disorder potential.  TTG is therefore unique among moir\'e engineered materials in that varying $V_\text{g}$ provides a means to systematically tune electronic disorder, thereby toggling between clean and dirty limits.  The interpretation of transport measurements in light of this behavior can lead to significant insights concerning the nature of the superconducting phase.  In particular, the co-occurrence of the flat band resonance condition, with its resulting minimization of electronic disorder, and optimal doping for superconductivity strongly suggests that the superconducting phase boundary along the doping axis is disorder driven.  This has certain implications for the symmetry of the superconducting order parameter \cite{TripletPairingClassification}.

In conventional s-wave superconductors the order parameter is robust against the presence of non-magnetic impurities, a result known as Anderson’s theorem \cite{Anderson1959}.  However, if the impurity potential is sufficiently dense on the scale of the  coherence length that direct superconducting pathways no longer percolate through the material, then superconducting correlations may develop locally even while disorder obstructs the formation of global phase coherence.  In this dirty limit, the system is governed by two energies: one associated with the formation of Cooper pairs within superconducting regions, and the other with the disorder induced barrier to tunneling between regions.  This behavior can be detected in transport as a two-step transition in the R-T curve \cite{Eley2012}.  

At zero magnetic field the superconducting transition in TTG exhibits a single sharp drop in resistivity\cite{pjh,kim}, implying that adjacent magic moir\'e grains are well coupled.  In this scenario, the suppression of superconductivity at dopings away from flat band resonance would likely be driven by pair-breaking twiston scattering, consistent with an unconventional pairing mechanism and nodal gap function.  The development of a second transition, or \textquotedblleft{}knee\textquotedblright{}\cite{pjh}, in the R-T curve at finite magnetic field, on the other hand, may be associated with the threading of flux through variously sized grains of magic moir\'e, resulting in global phase decoherence at the temperature at which Cooper pairs are initially bound.  Future work that systematically explores this expanded phase space by controllably tuning moir\'e defect density through the angle mismatch $\delta_\theta$ has the potential to shed further light on the pairing mechanism in TTG by determining its sensitivity to non-magnetic impurity scattering, as has been done in a range of other unconventional systems \cite{PhysRevLett.80.161,PhysRevB.85.214509,cuprate_impurities}.

\medskip
\bibliography{references}

\bibliographystyle{Science}
\medskip

\section*{Acknowledgements}
We thank Dorri Halbertal, Hector Ochoa and Alexei Tsvelik for fruitful discussions. Studies of the electronic structure of twisted trilayer graphene were supported as part of Programmable Quantum Materials, an Energy Frontier Research Center funded by the US Department of Energy (DOE), Office of Science, Basic Energy Sciences (BES), under award DE-SC0019443. The synthesis of the trilayer samples was supported by the NSF MRSEC program through Columbia in the Center for Precision-Assembled Quantum Materials (PAQM) - DMR-2011738. Support for cryogenic STM measurements was provided by the Air Force Office of Scientific Research via grant FA9550-16-1-0601. Z.Z and E.K are supported by STC Center for Integrated
Quantum Materials, NSF Grant No. DMR-1231319, ARO MURI Grant No. W911NF14-0247,
and NSF DMREF Grant No. 1922165. Relaxation calculations were performed on the Odyssey cluster supported by the FAS Division of Science, Research Computing Group at Harvard University.  M.C. and S.S. are supported by NSF Grant No. DMR-2002850.  K.W. and T.T. acknowledge support from the Elemental Strategy Initiative conducted by the MEXT, Japan (Grant Number JPMXP0112101001) and JSPS KAKENHI (Grant Numbers 19H05790 and JP20H00354).

\section*{Competing Interests}
The authors declare no competing interests.

\section*{Supplementary Materials}
Materials and Methods\\
Supplementary Text\\
Figs. S1 to S10\\

\newgeometry{left=0.625in,bottom=1in}
\newpage
\begin{figure}[htp]
    \centering
    \includegraphics[width=7.25in]{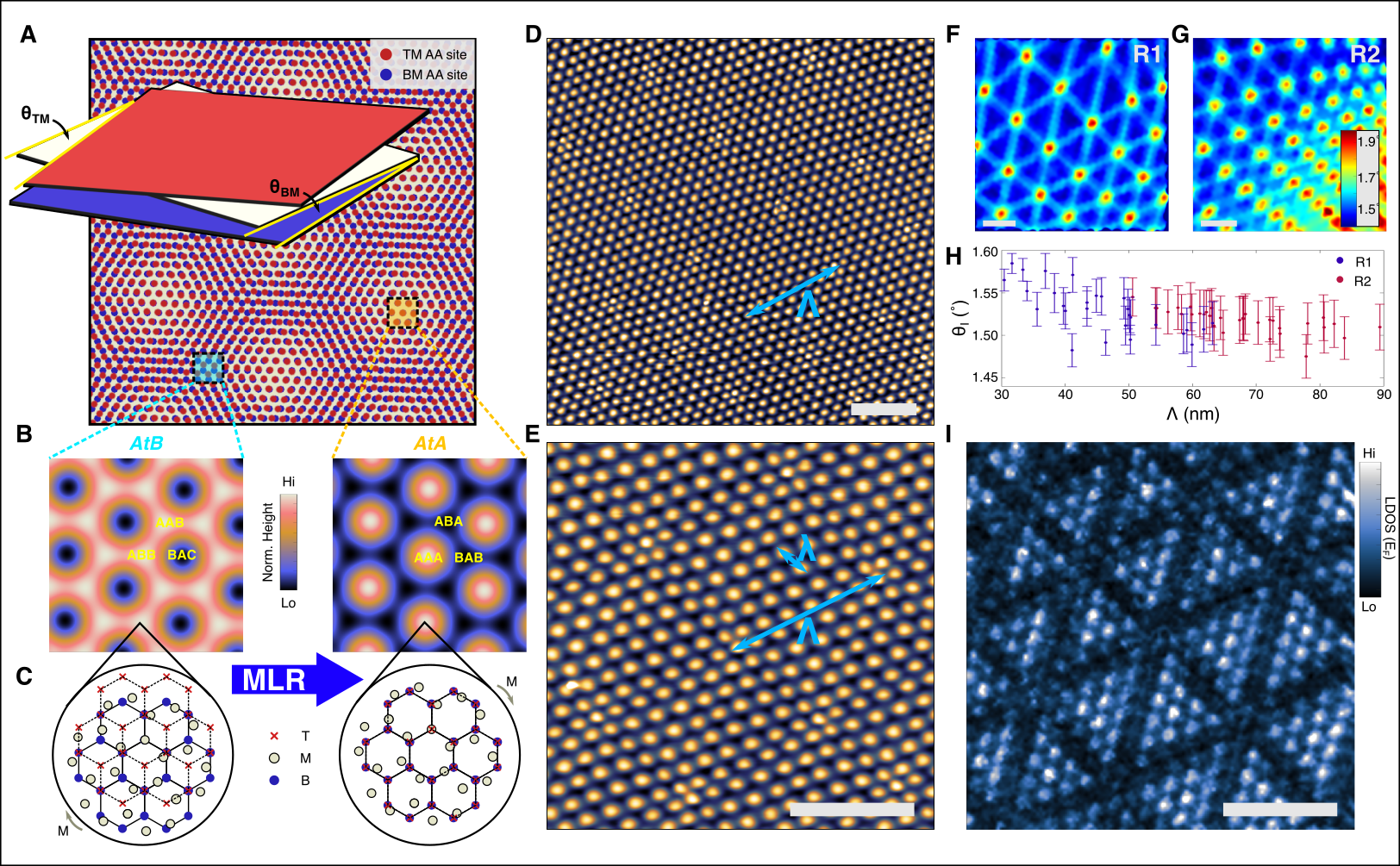}
    \caption{\textbf{STM on three twisted graphene layers.}  (\textbf{A})~Illustration of the moir\'e of moir\'e pattern in TTG for $\theta_{TM}\neq\theta_{BM}$ in the absence of lattice relaxation.  Local AtA and AtB domains are formed, creating two characteristic length scales.  Inset: Illustration of the two independent twist angles expected in a general three layer stack.  (\textbf{B})~Normalized out of plane corrugation calculated for AtB and AtA stacking configurations showing the local domain structure of each configuration.  (\textbf{C})~Schematic of the atomic stacking structure of AtA and AtB TTG.  The two configurations are related by translation of the top layer.  In real devices a moir\'e lattice reconstruction (MLR) makes it energetically favorable for AtB domains to warp into AtA.  (\textbf{D})~STM topography of TTG at an average twist of $1.56^\circ$.  (\textbf{E})~Zoomed in topography of a single unit cell of the larger moir\'e pattern in (D).  $\Lambda$ and $\lambda$ indicate two distinct moir\'e wavelengths.  (\textbf{F} and \textbf{G})~Local twist angle maps over two nearby sample areas.  The local twist angle is extracted from the cell areas of the Voronoi tessellation generated by the AAA site positions.  (\textbf{H})~Plot of the internal twist angle ($\theta_{I}$) within a MLR domain as a function of domain size ($\Lambda$) for regions R1 (F) and R2 (G).  Error bars represent one standard deviation of the local twist angle within a given domain.  (\textbf{I})~Charge neutral local density of states map acquired at the Fermi level showing electronic inhomogeneity due to the MLR.  $\text{V}_\text{set}$ = 300~mV, $\text{I}_\text{set}$ = 120 pA, $\text{V}_\text{mod}$ = 2~mV.  All scale bars are 50 nm.}
\end{figure}

\newpage
\begin{figure}[htp]
    \centering
    \includegraphics[width=7.25in]{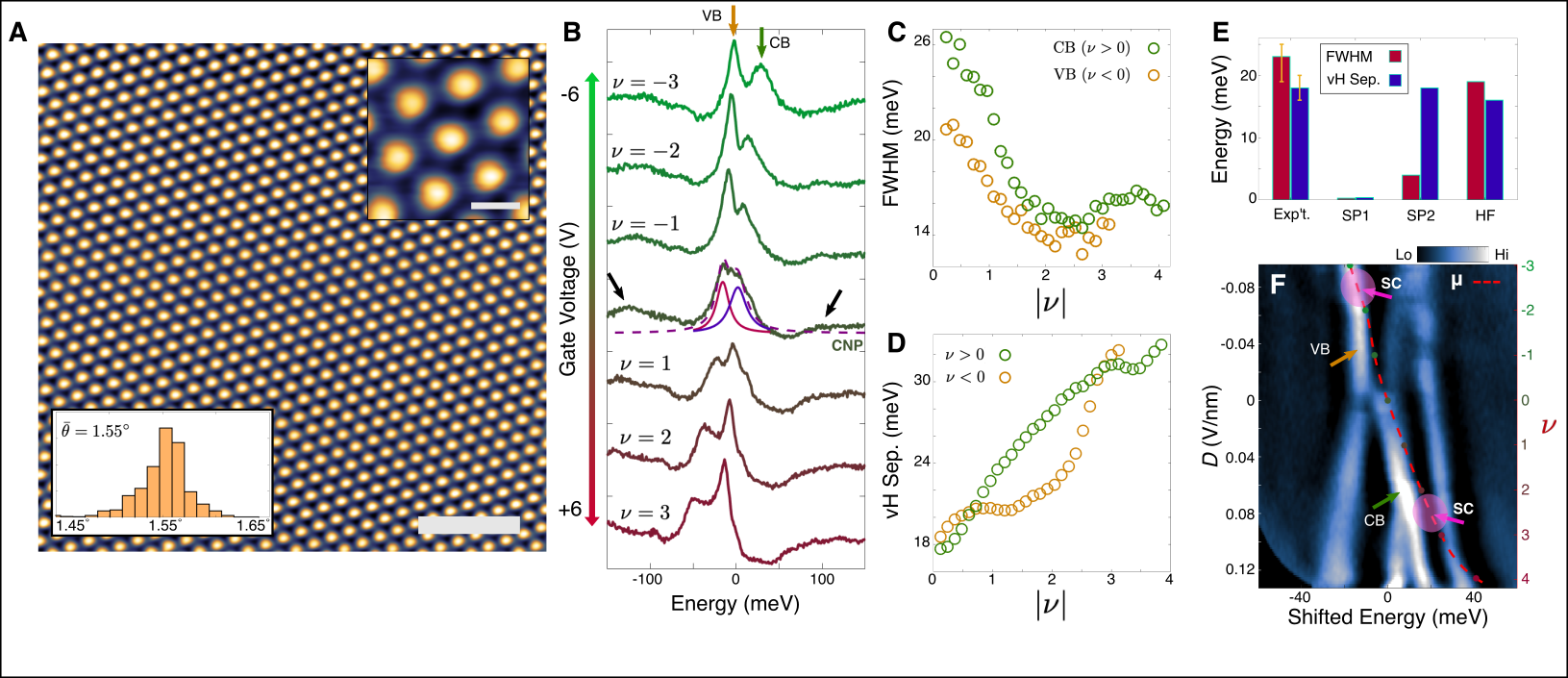}
    \caption{\textbf{Spectroscopy on a uniform 1.55$^\circ$ region.}  (\textbf{A})~STM topography of a uniform area presenting a single moir\'e wavelength corresponding to a twist angle of $1.55^\circ$.  Scale bar 50 nm.  Insets: Zoomed in topography of a single moir\'e unit cell showing bright AAA sites surrounded by alternating ABA and BAB domains (scale bar 8 nm); histogram of local twist angle values extracted for each moir\'e unit cell as in Figs.~1E and 1F.  (\textbf{B})~AAA site STS spectra showing the evolution of the flat band structure at $1.55^\circ$ twist as a function of applied gate voltage.  Each curve represents the average of 10 measurements performed on a single AAA site from the region in (A).  Gold and green arrows indicate the valence and conduction flat bands respectively.  Black arrows indicate the edges of the remote bands.  Charge neutral spectrum shows Lorentzian fits to the valence and conduction bands.  $\text{V}_\text{set}$ = 300~mV, $\text{I}_\text{set}$ = 150 pA, $\text{V}_\text{mod}$ = 1~mV.  (\textbf{C})~Full width at half maximum of the conduction and valence band VHSs from (B) as a function of respective band filling, showing that each band grows flatter as it is doped to the Fermi level.  (\textbf{D})~Separation between conduction and valence band peaks from (B) as a function of doping.  (\textbf{E})~Comparison of VHS separation and widths at charge neutrality between experiment and three continuum model calculations.  SP1 and SP2 are single particle calculations with different inter- and intra-layer hopping parameters, and HF includes electronic interactions \textit{via} Hartree-Fock corrections to the continuum model, resulting in a band renormalization, and lifetime broadening (see SM for details). Only the interacting calculation reproduces the experimental spectrum.  (\textbf{F})~High resolution AAA site spectra shifted as described in the text and with a smooth background subtracted to emphasize the evolution of the flat bands with doping.  Red dashed line denotes the position of the chemical potential for which a given spectrum was acquired.  Pink arrows indicate optimal doping for superconductivity.  $\text{V}_\text{set}$ = 200~mV, $\text{I}_\text{set}$ = 200 pA, $\text{V}_\text{mod}$ = 0.5~mV.}
\end{figure}

\newpage
\begin{figure}[htp]
    \centering
    \includegraphics[width=4.75in]{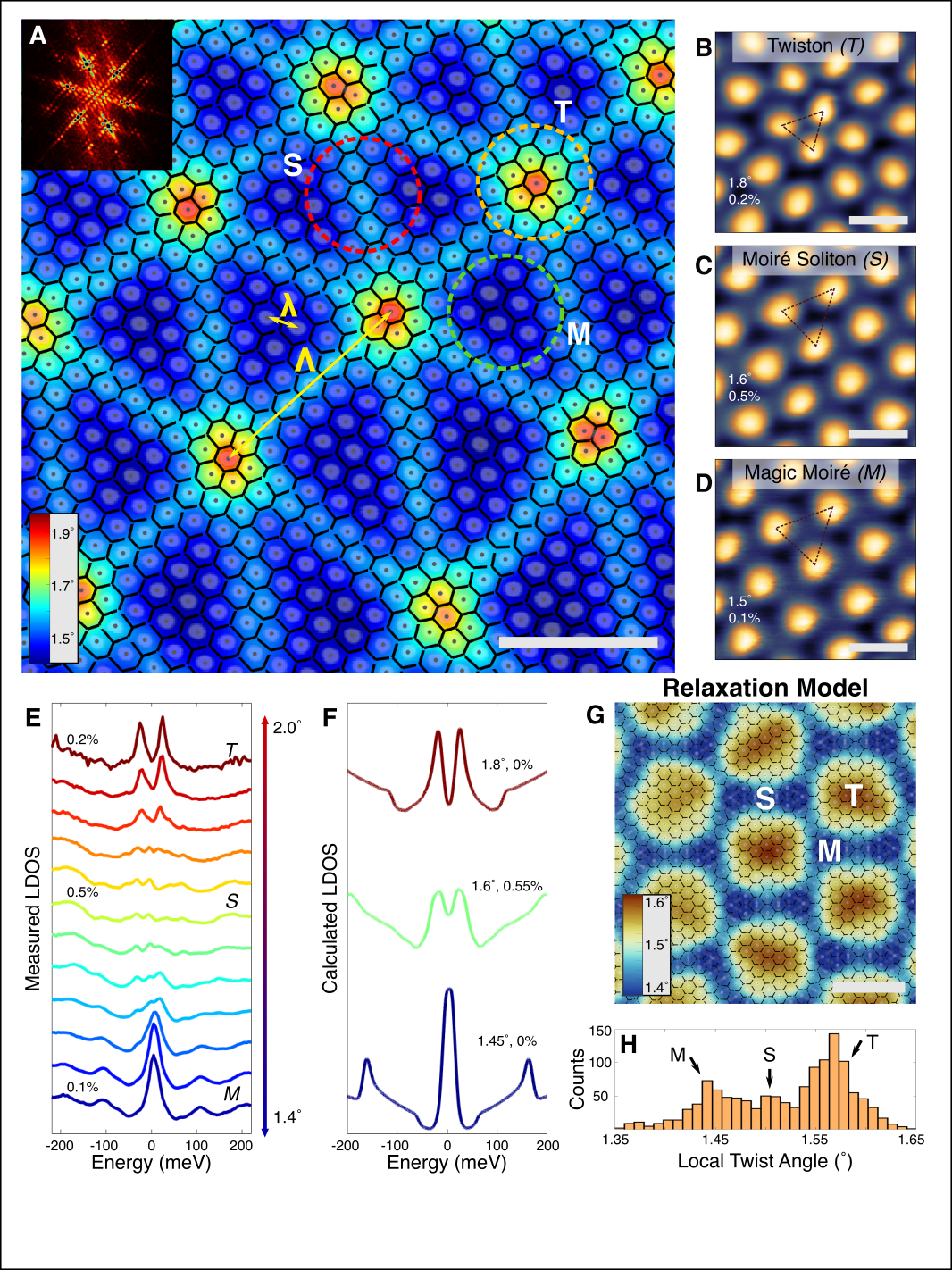}
    \caption{\textbf{Moir\'e lattice reconstruction.}  (\textbf{A})~STM topography colored in proportion to the local twist angle. Scale bar 50 nm.  Inset: FFT of 320 nm$^{2}$ topograph centered on this field of view showing two sets of moir\'e wave vectors.  (\textbf{B} - \textbf{D})~Zoomed in topography of the circled regions in panel (A) illustrating the local structure of the MLR.  Numerical insets indicate local twist angle and heterostrain values extracted from dashed moir\'e lattice vectors.  Scale bars 10 nm.  (\textbf{E})~Experimental AAA site LDOS spectra extracted from conductance maps taken over the field of view of panel (A) displaying the change in electronic structure over different regions of the MLR.  Percentages denote characteristic heterostrain values for each MLR region.  (\textbf{F})~Continuum model (SP2) TTG densities of states for three sets of structural parameters ($\theta,\varepsilon$).  Calculations at finite heterostrain preserve mirror symmetry by applying a uniaxial strain to the middle layer only.  (\textbf{G})~Local twist angle as determined by nearest neighbor AAA site distance for structural relaxation calculation with $\theta_{TM} = 1.5^\circ$ and $\theta_{BM} = 1.69^\circ$.  Scale bar 50 nm.  (\textbf{H})~Histogram of the twist angles present in panel (G) showing three populations corresponding to magic, soliton, and twiston sites.  }
\end{figure}

\newpage
\begin{figure}[htp]
    \centering
    \includegraphics[width=7.25in]{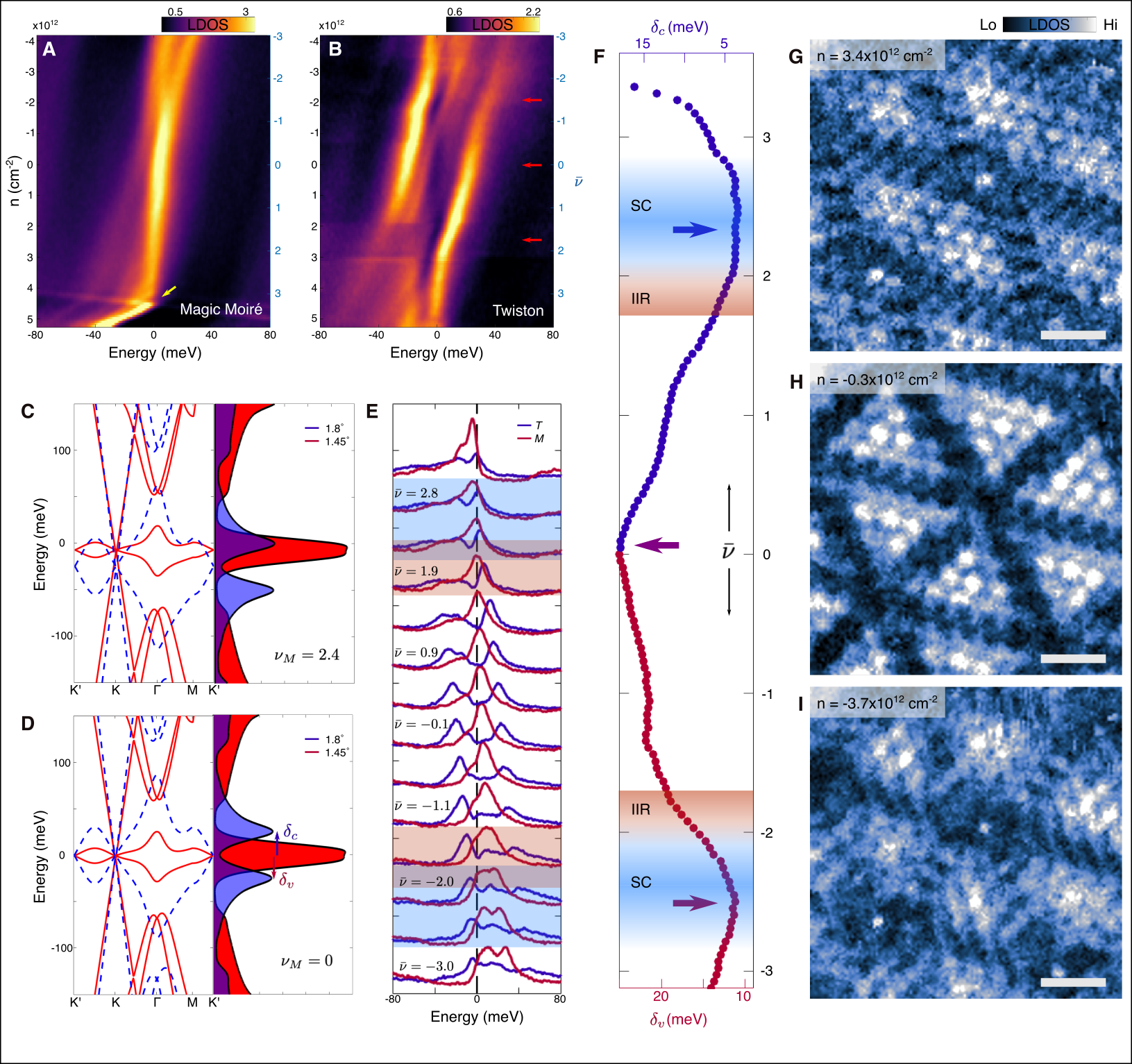}
    \caption{\textbf{Twiston gaps and flat band resonance.}  (\textbf{A} and \textbf{B})~Gate dependent LDOS spectroscopy on the magic moir\'e and twiston regions.  Yellow arrow in (A) indicates full filling of the moir\'e superlattice.  Red arrows in (B) indicate correlated gaps that are confined to the twiston region.   (\textbf{C} and \textbf{D})~Continuum model calculations showing the band structure (left) and density of states (right) for twist angles of 1.45$^\circ$ (red) and 1.8$^\circ$ (blue) at two different fillings.  The zero of energy corresponds to the position of the chemical potential.  Flat band resonance (C) for these angles occurs when the magic moir\'e (1.45$^\circ$) superlattice is filled to $\nu_{M} = 2.4$.  (\textbf{E})~Doping dependent LDOS spectroscopy on the twiston (blue) and magic moir\'e (red) regions showing flat band resonance at $|\bar{\nu}|\sim2.5$. Vertical dashed line represents the Fermi level.  (\textbf{F})~Extracted values of flat band energy splitting between twiston and magic sites, $\delta_{c/v}$ (see panel (D)), as a function of doping.  Minima correspond to flat band resonances and the resulting reduction in real space electronic disorder.  (\textbf{G}-\textbf{I})~Fermi level LDOS maps at charge neutrality (H) and at the electron (G) and hole (I) doped flat band resonances.  Scale bars 25 nm.  $\text{V}_\text{set}$ = 300~mV, $\text{I}_\text{set}$ = 120 pA, $\text{V}_\text{mod}$ = 2~mV.}
\end{figure}

\restoregeometry

\end{document}